\documentclass[final]{svjour2}
\usepackage{graphicx}
\usepackage{rotating}
\usepackage{amssymb}
\usepackage{mathptmx}
\usepackage[numbers]{natbib}
\makeatletter
\journalname{Journal of Low Temperature Physics}
\bibpunct{}{}{,}{s}{}{,}

\begin{document}

\newcommand{\hdblarrow}{H\makebox[0.9ex][l]{$\downdownarrows$}-}
\title{Vibration Isolation Design for the Micro-X Rocket Payload}

\author{S.N.T. Heine \and E. Figueroa-Feliciano \and J.M. Rutherford \and P. Wikus \and P. Oakley \and F.S. Porter \and D. McCammon}

\institute{MIT Kavli Institute, Massachusetts Institute of Technology,\\ Cambridge, MA 02139, USA\\
\email{saraht@mit.edu}}

\date{10.12.2013}

\maketitle

\begin{abstract}

Micro-X is a NASA-funded, sounding rocket-borne X-ray imaging spectrometer that will allow high precision measurements of velocity structure, ionization state and elemental composition of extended astrophysical systems.  One of the biggest challenges in payload design is to maintain the temperature of the detectors during launch.  There are several vibration damping stages to prevent energy transmission from the rocket skin to the detector stage, which causes heating during launch.  Each stage should be more rigid than the outer stages to achieve vibrational isolation.  We describe a major design effort to tune the resonance frequencies of these vibration isolation stages to reduce heating problems prior to the projected launch in the summer of 2014.


\end{abstract}

\section{Introduction}
The Micro-X focal plane consists of an array of transition edge sensor (TES) microcalorimeters, each of which measures the temperature change of an absorber due to an incident photon in the soft X-ray band with projected resolution of 2--4~eV.  In order to maintain high energy resolution, the detectors are cooled to 50 mK by an adiabatic demagnetization refrigerator (ADR) to limit thermal noise \cite{detectors}.    The duration of the sounding rocket flight will be approximately 15 minutes, with the observation taking about 5 minutes, so recycling of the ADR in flight is not possible \cite{MicroXdesign}.  Therefore, the ADR must be cycled before flight and the detectors kept cold during launch.  This requires a very high level of vibration isolation between the skin and the detector stage to shield the detectors from heating caused by the near-white, 12.7 g$_{rms}$ vibration spectrum of the Terrier Black Brandt rocket launch \cite{rockethandbook}.  

The vibration isolation system for the Micro-X payload has several stages: damping suspension pieces between the skin and the dewar, a system of G10 tubes inside the dewar supporting the helium can and a kevlar suspension to isolate the detector stage from the liquid helium stage \cite{MicroXdesign}.  Isolating the inner stages from the vibration at the outer stages is achieved by staggering the resonant frequencies of the stages in such a way that the resonant frequency of the stages increase closer to the detector.  This means that the coldest stages are the stiffest, and have resonant frequencies that sit in regions that are damped by each of the stages acting between the cold stage and the skin \cite{PatrickICC}.  In June, 2012, we travelled to Wallops flight facility to submit the payload to launch-level vibration testing.  This testing resulted in detector stage heating far above acceptable limits and prompted a testing and redesign phase discussed below. 

\begin{figure}
\begin{center}
\includegraphics[%
  width=0.75\linewidth,
  keepaspectratio]{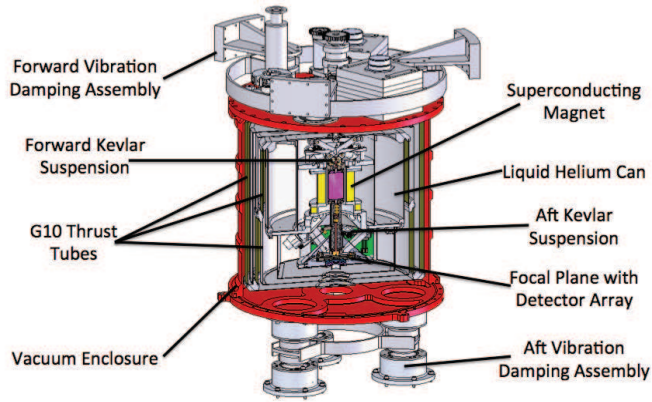}
\end{center}
\caption{(Color online) A schematic of the Micro-X dewar and dewar suspension.  In this schematic the forward side of the dewar is up and the aft side is down.  The pieces marked "damping assembly" connect directly to the rocket skin on the forward side and the bulkhead (also rigidly connected to the rocket skin) on the aft side.}
\label{fig: schematic}
\end{figure}

\section{Lab vibration setup}
\label{sec: setup}
\begin{figure}
\begin{center}
\includegraphics[%
  width=0.55\linewidth,
  keepaspectratio]{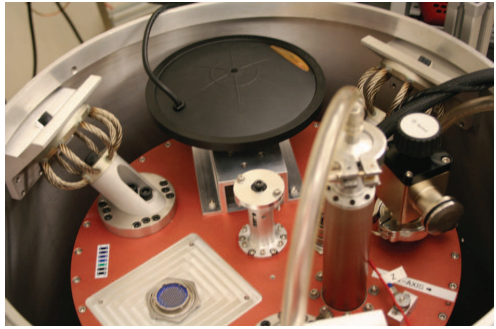}
\end{center}
\caption{(Color online) A picture of the in-lab vibration setup.  The red can is the Micro-X cryostat, mounted to the flight skin-the aluminum cylinder around it.  The transducer (Clark Synthesis TST329 Gold Transducer-the black disk mounted on the can near the top of the image) can be mounted on adapter pieces to provide vibration along any axis.}
\label{fig: vibesetup}
\end{figure}

Since trips to a vibration table are costly, we created an inexpensive, in-house vibration setup to test design iterations using an audio transducers sold commercially as home theater equipment.  The transducer mounts to the top plate of the dewar so that we can vibrate the dewar directly.  We drive the transducer with a slow sine sweep produced by a dynamic signal analyzer, monitoring either accelerometers at different stages of the dewar at room temperature or the heating produced by the sweep when the liquid helium and detector stages are at 4 K.  We have found that the accelerometer data is often difficult to interpret, so our most reliable method has been to use the heating data.  This method does not take into account the frequency response of the transducers themselves, but has been quite reliable in helping us to understand changes between design iterations.  A picture of the setup with the audio transducer mounted to the top plate of the dewar in the Z (thrust) direction is shown in Figure \ref{fig: vibesetup}.

\section{Kevlar Suspension System Redesign}
\begin{figure}
\begin{center}
\includegraphics[%
  width=0.61 \linewidth,
  keepaspectratio]{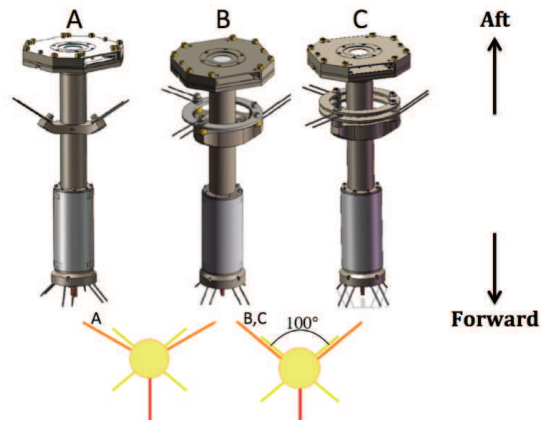}
\end{center}
\caption{Schematics of the suspended mass for the three design iterations of the suspension.  Design A was our initial design tested in the June 2012 vibration test, design B was an intermediate iteration and design C is our current design that was subjected to a launch-level vibration test in August 2013.  Side views are shown above while top views showing the angular orientation of the strings is shown below with forward strings shown in yellow, aft non-tensioning strings shown in orange and the tensioning string shown in red.  Four of the seven kevlar attachment points are on the forward side of the stage and the remaining three including the tensioning piece are on the aft end of the stage.}
\label{fig: Suspdesigns}
\end{figure}

\begin{figure}
\begin{center}
\includegraphics[%
  width=0.83\linewidth,
  keepaspectratio]{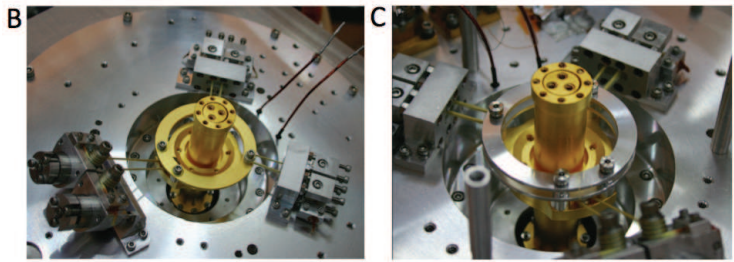}
\end{center}
\caption{(Color online) Pictures of the aft side of the suspension without the detector box installed for designs B (left) and C (right).  Each kevlar attachment consists of a single piece of kevlar glued down on both ends and wrapped around a spiral part at the liquid helium stage and looped around a cylindrical attachment on the detector stage.  This can be most clearly seen in these pictures on the tensioning piece (leftmost) in the design B picture.}
\label{fig: Susppics}
\end{figure}

Our largest redesign effort has been in the kevlar suspension system for the detector stage.  The system is made up of seven kevlar attachments: six to constrain the position of the stage and one containing springs for tensioning.    See Figure \ref{fig: Suspdesigns} for schematics of this system and Figure \ref{fig: Susppics} for pictures of the aft side of the system.

%\begin{figure}
%\begin{center}
%\includegraphics[%
  %width=0.45\linewidth,
  %keepaspectratio]{SuspTopViews}
%\end{center}
%\caption{(Color online) Top views of the orientation of the seven kevlar strings.  The yellow lines represent the strings on the forward side, the orange represent the non-tensioning strings on the aft side and the red represents the tensioning piece.  For designs B and C the non-tensioning aft strings were moved closer together to strengthen the suspension in the direction opposite the tensioning string.}
%\label{fig: Susptopview}
%\end{figure}

At our vibration test in June 2012, we found our initial design (design A in Figure \ref{fig: Suspdesigns}) exhibited a resonance frequency near 200 Hz, which is significantly lower than was predicted, and very near that of the helium stage.  The coupling between these resonances led to a large amplitude of vibration of the detector stage causing excessive heating.  After additional FEM modeling we determined that the non-tensioning aft strings being oriented at an angle between the vertical and horizontal axes led to coupling between the axes, and encouraged a bowling-pin-like mode of the detector.  Our first redesign (shown as design B in Figure \ref{fig: Suspdesigns}) oriented these pieces horizontally (normal to the axis of the rocket) and moved them closer together so that the component of their force opposite the tensioning piece on the aft side was larger, strengthening the suspension in that direction.  See Figure \ref{fig: Suspdesigns} for schematics illustrating these design changes.  Design B exhibited a broad resonance peak around 285 Hz.  Despite this, we still saw strong heating around 200 Hz when we excited the dewar using a sine sweep in our lab setup as described in Section \ref{sec: setup}.  We believe this was due to coupling between the helium stage resonance at 200 Hz and the side wing of the broad detector stage resonance at 285 Hz.

These results prompted an additional redesign effort producing the design labelled C in Figure \ref{fig: Suspdesigns}.  Pictures of designs B and C are shown in Figure \ref{fig: Susppics}.  We determined that bending of the screws at the attachment point of the kevlar on the detector stage could both lower the resonance frequency of the entire system and account for some of the broad nature of the resonance we witnessed.  We therefore added a stabilizing ring at the top of these screws to keep them from bending.  In addition, we used thicker kevlar (8520 denier rather than 4650 denier) on the two horizontal strings on the aft side to provide additional strengthening against horizontal motion.  Finally, we determined that the spirals, which are designed to provide frictional isolation for the glue joint from the full tension on the kevlar (shown in Figure \ref{fig: Susppics} ) were still sufficiently able to provide this isolationt with the string wrapping around them for only a quarter of a turn rather than the three and a quarter turns we were initially using.  This helped to stiffen the system because as a string turns around a radius the innermost strands are preferentially engaged, so the less radius the string travels around, the more strands of kevlar participate in stiffening the system.  

\begin{figure}
\begin{center}
\includegraphics[%
  width=0.55\linewidth,
  keepaspectratio]{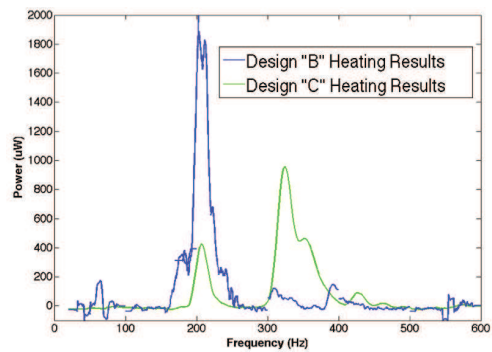}
\end{center}
\caption{(Color online) Heating results due to vibration of the dewar with the detector stage at 4 K.  Results from tests of design B are shown in blue and have a main peak at 200 Hz.  Results from tests of design C are shown in green and have a much smaller peak at 200 Hz with a main peak at 325 Hz.  We convert temperature slope to power, by measuring the temperature slope produced by known applied power using a heater on the detector stage.}
\label{fig: heating}
\end{figure}

These changes led to a shift in resonance frequency from 285 Hz to about 325 Hz, along with a reduction in the width of the resonance peak.  When tested in our lab system we now see a much smaller heating peak around 200 Hz and a new peak at 325 Hz showing we have effectively decoupled the resonances of the helium tank and the suspended mass.  The data from the heating tests of designs B and C is shown in Figure \ref{fig: heating}.  

\begin{figure}
\begin{center}
\includegraphics[%
  width=0.45\linewidth,
  keepaspectratio]{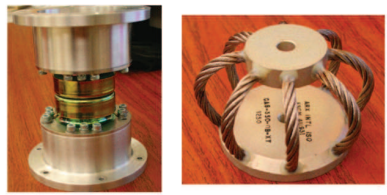}
\end{center}
\caption{(Color online) Rubber dampers from Barry Control with attachment fixtures (left) and wire rope isolator from Isotech, Inc. (right).  The wire rope isolators are shown suspending the cryostat in Figure \ref{fig: vibesetup}}
\label{fig: dampers}
\end{figure}
        
\begin{figure}
\begin{center}
\includegraphics[%
  width=0.60\linewidth,
  keepaspectratio]{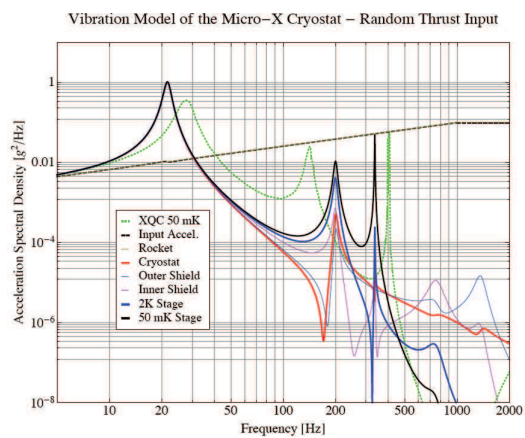}
\end{center}
\caption{(Color online) In addition to the CAD modeling which provides the expected normal modes of the system, we have modeled the system with a six degree of freedom, one dimensional model in Mathematica to predict the motion of each stage of our cryostat, which provides a rough expectation for the coupling and heating expected from the full system.  Inputs are the spring constants between each stage and the spectrum of excitation.  The vibration spectrum of the X-ray Quantum Calorimeter (XQC) rocket on which our design is based is shown for comparison \cite{XQC}.}
\label{fig: model}
\end{figure}

\section{Next Steps}
Another targeted area for redesign is the system suspending the dewar from the rocket skin.  We initially used rubber dampers  produced by Barry Control to suspend the dewar from the skin.  In our initial vibration test, the dewar showed a broad resonance in the 40 to 50 Hz range when suspended with the rubber dampers.  Since then we have transitioned to a system of wire rope isolators from Isotech, Inc., which showed a resonance around 30 Hz in our vibration test in August, 2013.  Each of these dampers is shown in Figure \ref{fig: dampers}.  The test in August showed acceptable heating levels in full launch load vibrations on both lateral axes and a much reduced but still high heating level during a full launch load vibration in the thrust (vertical) axis.  We have identified the cause of this heating as a coupling between a resonance in the skin system and the helium tank resonance around 200 Hz and are currently designing a modification in the G10 suspension system to move the helium tank resonance down.  We will test this modification at Wallops in late 2013 and move forward towards integration and testing for a summer 2014 launch.


\begin{thebibliography}{99}

\bibitem{detectors}
Bandler S. R. et al., {\it J. of Low Temp. Phys.}, \textbf{151}, 400-405, (2008).

\bibitem{rockethandbook}
Goddard Space Flight Center, Wallops Flight Facility, {\it NASA Sounding Rocket Program Handbook}, (2005).

\bibitem{MicroXdesign}
P. Wikus, J. S. Adams, Y. Bagdasarova, S. R. Bandler, W. B. Doriese, M. E. Eckart, E. Figueroa-Feliciano, R. L.
Kelley, C. A. Kilbourne, S. W. Leman, D. McCammon, F. S. Porter, J. M. Rutherford and S. N. Trowbridge, {\it Advances in Cryogenic Engineering} \textbf{55}, 633, (2010).

\bibitem{PatrickICC}
P. Wikus, J.M. Rutherford, S.N. Trowbridge, D. McCammon, J.S. Adams, S.R. Bandler, R. Das, W.B. Doriese, M.E. Eckart, E. Figueroa-Feliciano, R.:. Kelley, C.A. Kilbourne, S.W. Leman, F.S. Porter and K. Sato, {\it Proceedings of ICC} (2010).

\bibitem{XQC}
D. McCammon et. al., {\it Nucl. Instrum. Methods Phys. Res.,} Sect. A \textbf{370}, 266-268, (1996).

\end{thebibliography}
\end{document}